\documentclass{PoS}
\title{$\Delta\eta\Delta\phi$ angular correlations in pp collisions at the LHC registered by the ALICE experiment}

\ShortTitle{$\Delta\eta\Delta\phi$ correlations in ALICE}

\author{\speaker{Ma{\l}gorzata Janik} (for the ALICE Collaboration)\\
        Warsaw University of Technology, Poland\\
        E-mail: \email{majanik@if.pw.edu.pl}}

\abstract{We report on studies of two-particle $\Delta\eta\Delta\phi$ angular correlations
measured in proton-proton collisions at center of mass energies
$\sqrt{s} = 0.9\ \rm{TeV}$, $\sqrt{s} = 2.76\ \rm{TeV}$  and $\sqrt{s} = 7\ \rm{TeV}$ registered by the ALICE
experiment at LHC. We present the dependence of the correlation function on the pair
transverse momentum, the multiplicity of the event and the charge
combination of particles in the pair. $\Delta\eta\Delta\phi$ correlations are
expected to exhibit several structures which arise from different
physics mechanisms. The results are also related to the correlations obtained by femtoscopic
analysis. We show that the minijet hypothesis for the non-femtoscopic
underlying correlation should be taken into account while performing
the femtoscopy analysis.}

\FullConference{The Seventh Workshop on Particle Correlations and Femtoscopy\\
		 September 20 - 24 2011\\
		 University of Tokyo, Japan}

\begin{document}

\section{Introduction}
In high energy collisions different mechanisms come into play during the dynamic evolution of the system of particles. The observed features in the final shape of the system results from the complex interplay of these processes. Studies of such physical mechanisms in the experimental data require the development of special analysis tools. One such tool is the analysis of the angular correlations in $\Delta\eta\Delta\phi$ space, where different correlation sources contribute to the final result. 

We report results of the $\Delta\eta\Delta\phi$ angular correlations in pp collisions at 
$\sqrt{s} = 0.9$, $2.76$ and $7$~TeV registered by the ALICE experiment \cite{ALICE}. 
We preform the $\Delta\eta\Delta\phi$ angular correlations analysis distinguishing between correlations for like-sign and unlike-sign particle pairs.
The main goal of the performed analysis is to characterize two of many correlation sources that contribute to the $\Delta\eta\Delta\phi$ correlation function:  minijets (particles from fragmentation of "hard scattered" partons, which have too low energy in comparison to the underlying event to be reconstructed as jets event-by-event) and femtoscopic correlations (femtoscopy is a term which refers to the technique of two-particle nuclear interferometry, which combines the momentum distributions of particles with the space-time characteristics of their sources). We studied the dependence of the correlation function on particle charge, multiplicity, and pair transverse momentum to disentangle those two correlation sources and study their properties separately. Moreover, we tested the hypothesis of minijets being the background for the femtoscopic correlation functions discussed in \cite{NewFemtoPaper}.

\section{Data Analysis}
\subsection{Event and track selection}
The analyzed datasets were 11.3 millions events at $0.9\ \rm{TeV}$, 12.6 milion events at $2.76\ \rm{TeV}$ and 530 millions at $7\ \rm{TeV}$.
Minimum bias data samples were used for the analysis. Events with the primary vertex within $1\ \rm{mm}$ of the beam in the transverse plane and within $10\ \rm{cm}$ of the center of the ALICE detector along the beam axis were selected.
The analysis was performed on primary particles within the acceptance range of       
$|\eta|<1.0$  and the transverse momentum range $p_T>0.12$ GeV/c. Moreover, cuts to remove effects of splitting (one track is mistakenly reconstructed as two) and merging (two tracks are reconstructed as one or not at all) were applied.

We applied a cut to remove electron-positron pairs which are produced from gamma conversions. It was achieved by removing pairs constructed of one positive and one negative particle that have small $\Delta\theta$ polar angle difference and invariant mass close to $0$ (rest mass of the photon).

We observe a combination of undesired physics effects (gamma conversions), true physics correlations (Coulomb interaction) and detector effects in the (0,0) bin. Their systematic study is in progress. At the moment we assign a systematic uncertainty for this bin of 0.1 (absolute with respect to the correlation strength) and 2\% (relative to the correlation strength) for all other bins.

To compare with the collision data, the same analysis was performed on Monte Carlo events generated by two different models -- Pythia and Phojet.

\subsection{Correlation function}
\label{chap:results}

The experimental correlation function is defined as the ratio of the
signal to the background. The two-particle correlation as a function
of $\Delta\eta$ and $\Delta\phi$ is defined as follows: 
\begin{equation}
\label{eq:CorrelationFuntion}
C(\Delta\eta,\Delta\phi)=\frac{N_{pairs}^{mixed}}{N_{pairs}^{signal}} \frac{S(\Delta\eta,\Delta\phi)}{B(\Delta\eta,\Delta\phi)},
\end{equation}
where $\Delta\eta=\eta_1 - \eta_2$ is the difference in
pseudorapidity, $\Delta\phi=\phi_1 - \phi_2$ is the difference in
azimuthal angle, $N_{pairs}^{signal}$ is the number of pairs
constructing the signal $S$, and $N_{pairs}^{mixed}$ is the number of
pairs in the background $B$. The signal is determined by counting particle pairs within the same
$\Delta\eta\Delta\phi$ range: 
\begin{equation}
\label{eq:Nsignal}
S(\Delta\eta,\Delta\phi)=\frac{d^2N^{signal}}{d\Delta\eta d\Delta\phi}.
\end{equation}
The background is estimated by applying the so-called event
  mixing 
 and can be expressed as:
\begin{equation}
\label{eq:Nmixed}
B(\Delta\eta,\Delta\phi)=\frac{d^2N^{mixed}}{d\Delta\eta\Delta\phi}.
\end{equation}

\subsection{Background of femtoscopic correlation function}
Non-flat background of the femtoscopic correlation functions was reported in \cite{NewFemtoPaper,Lukasz}. We try to assess with the $\Delta\eta\Delta\phi$ analysis if the hypothesis of minijets being responsible for this structure, as proposed in \cite{BackgroundHypotheis}, holds.  

$\Delta\eta\Delta\phi$ correlations are well-suited for this task, since there is a direct relation between $\Delta\eta$, $\Delta\phi$ and the components of relative momentum of the pair $\mathbf{q}$, which are used in femtoscopic studies: \begin{equation}
q_{out}\sim p_{T,1}-p_{T,2},
\end{equation}
\begin{equation}
q_{side}\sim (p_{T,1}+p_{T,2})\Delta\phi,
\end{equation}
\begin{equation}
q_{long}\sim (p_{T,1}+p_{T,2})\Delta\eta.
\label{eq:connection}
\end{equation}

This means that from studies of angular correlations in $\Delta\eta-\Delta\phi$ space it is possible to describe other, non-femtoscopic correlation sources, that influence the femtoscopic correlation functions.

\section{Results}

\subsection{Energy dependence}
\begin{figure}[!ht]
\centering
\includegraphics[width=15cm]{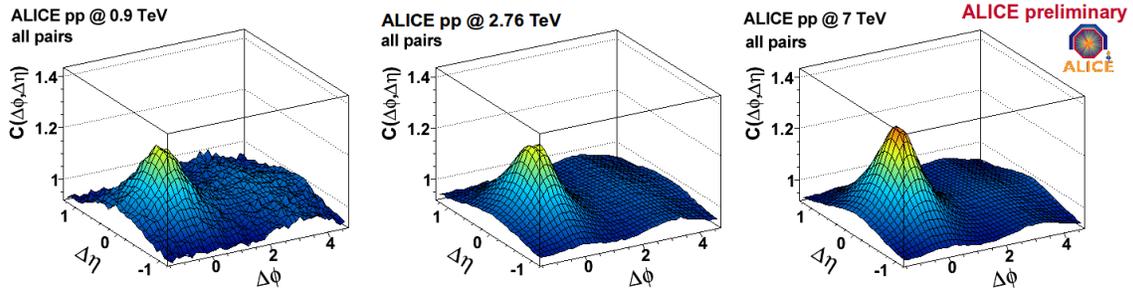}
\caption{\emph{Dependence of the $\Delta\eta\Delta\phi$ correlation functions on energy for pp collision data.}}
\label{fig:energy_dependence}
\end{figure}

Correlation functions for three different energies are shown in figure \ref{fig:energy_dependence}. For lower collision energy, we expect that there would be less hard-scattering processes (a smaller magnitude of the correlation function of minijets) which would result in smaller correlation. 

\subsection{Charge dependence}

\begin{figure}[!ht]
\centering
\includegraphics[width=14cm]{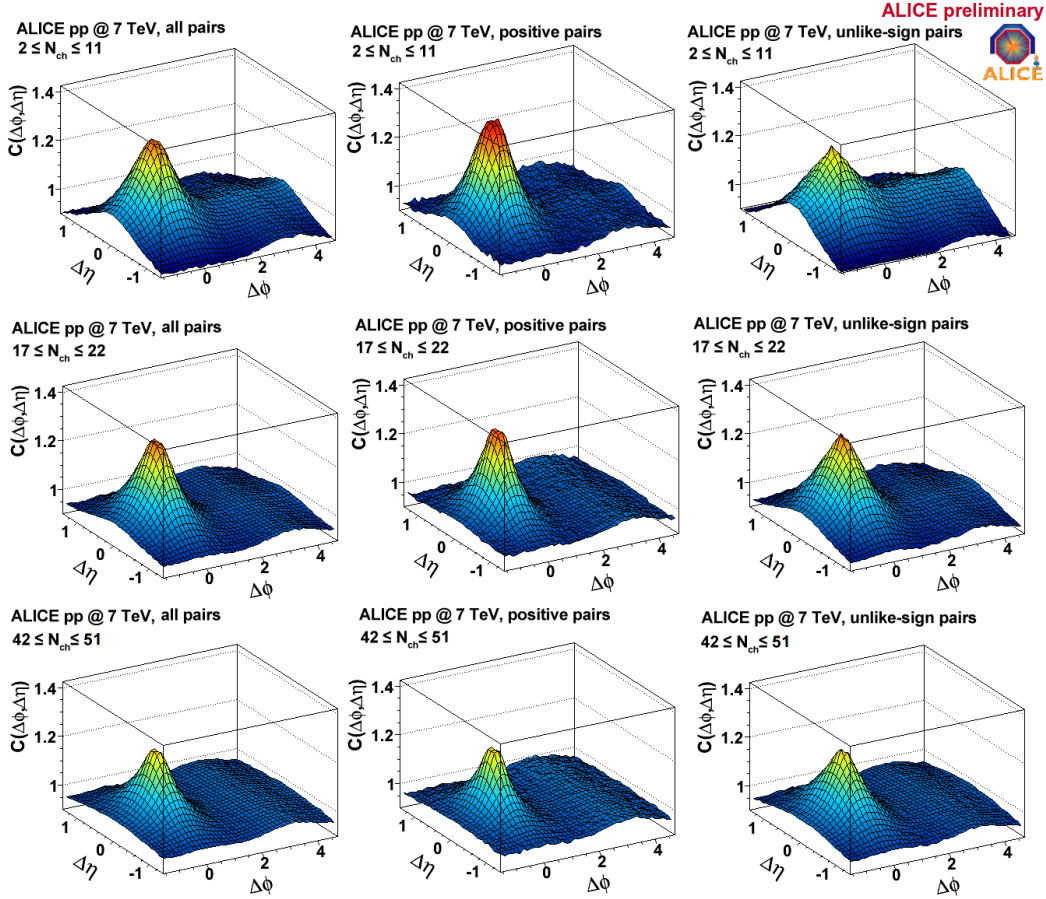}
\caption{\emph{Multiplicity dependence of $\Delta\eta\Delta\phi$ correlation function for $\sqrt{s}=7\ \rm{TeV}$ pp collision data.}}
\label{fig:multiplicity_dependence}
\end{figure}

Four different sets of particle pairs were analyzed: positive like-sign, negative like-sign, unlike-sign, and all possible (a pair can be combined of any particle, independently of charge). The last case is the usual way of studying $\Delta\eta\Delta\phi$ correlations. However, to be able to show the influence of minijets on the femtoscopic correlations, we must analyze separately like-sign and unlike-sign pairs. This is due to the fact, that Bose-Einstein correlations occur only for identical bosons.
The results of the charge dependent analysis performed for $\sqrt{s}=7\ \rm{TeV}$ pp collision data are shown, together with the multiplicity dependence, in figure \ref{fig:multiplicity_dependence}. As we can see the near-side peak at $(0,0)$, corresponding to the minijet and Bose-Einstein correlations, and the away-side ridge at $\Delta\phi=\pi$, corresponding to back-to-back minijets, are both present. Since the correlation functions for like-sign pairs combined of negative and positive particles are identical within the statistical uncertainties, only correlation functions for positive pairs are shown. There is a significant difference between like-sign and unlike-sign distributions. It is consistent with the assumption that there is an additional source of correlation of identical particle pairs due to the Bose-Einstein correlations which does not exist for unlike-sign pairs.

\subsection{Multiplicity dependence}
\label{etaphiresults:multidependence}
The events were divided into eight multiplicity ranges for the data from $\sqrt{s}=$~$7\ \rm{TeV}$ collisions. The multiplicity ranges have comparable number of pairs. The analysis was performed separately in each range, for different charge combinations of particles in a pair (all possible, like-sign, and unlike-sign). In figure \ref{fig:multiplicity_dependence} we compare three multiplicity bins for all particles, positive like-sign and unlike-sign.

Studies of femtoscopic correlations show that there is a decrease of the correlation with increasing multiplicity \cite{NewFemtoPaper}. Similar conclusions can be obtained from $\Delta\eta\Delta\phi$ studies by comparing the plots in figure \ref{fig:multiplicity_dependence}. This is also expected for minijet correlations: for high multiplicity events the number of minijets per collision increases; therefore different minijets become background for each other and the correlation per pair decreases. Similar conclusions can be obtained for $\sqrt{s}=0.9$ and $2.76\ \rm{TeV}$ pp collision data. 

The low multiplicity correlation functions for unlike-sign pairs show a prominent longitudinal ($\Delta\eta \sim 0$) ridge. This structure can be observed for low multiplicities, at low transverse momenta, for unlike-sign pairs and is not reproduced by Monte Carlo models well \cite{Presentation}.

\subsection{Pair Transverse Momentum dependence}
\begin{figure}[!ht]
\centering
\includegraphics[width=15cm]{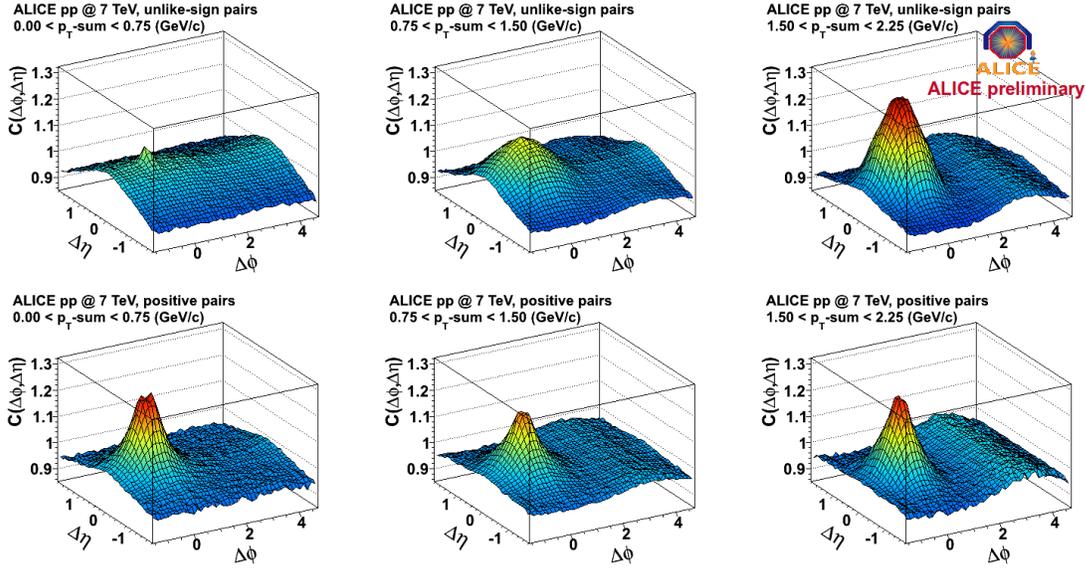}
\caption{\emph{Charge and $p_{T-sum}$ dependence of $\Delta\eta\Delta\phi$ correlation functions for $\sqrt{s}=7\ \rm{TeV}$ pp collision data.}}
\label{fig:pTsum_dependence}
\end{figure}

 Since femtoscopic correlations are most prominent for small pair transverse momenta, and minijets exactly the opposite, to disentangle the correlations coming from these two sources we introduced a new quantity which we call $p_T$-sum. It is defined by the equation:
\begin{equation}
p_{T-sum}=|\mathbf{p_{T,1}}|+|\mathbf{p_{T,2}}|.
\label{eq:ptsumdef}
\end{equation}
The formula in the equation (\ref{eq:ptsumdef}) includes the sum of the absolute values of the transverse momenta of the particles in the pair. We introduced four $p_T$-sum ranges: $0-0.75$, $0.75-1.5$, $1.5-2.25$, $2.25-10$ GeV/c. Results for the first three bins are shown in the figure \ref{fig:pTsum_dependence}.
In the upper panel of figure \ref{fig:pTsum_dependence} we show correlation functions for unlike-sign pairs. For the lowest $p_{T-sum}$ bin we observe that no near-side peak nor away-side ridge are present. When we move to higher transverse momenta those structures emerge. On the other hand, if we look at the plots of the lower panel (for like-sign pairs) we can see a prominent near-side peak for the lowest $p_{T-sum}$ bin, without any away-side ridge. It is consistent with our assumption, that for small pair transverse momenta the near-side peak is built mostly from Bose-Einstein correlations (no similar structure for unlike-sign, no away-side ridge). When we go to higher transverse momenta the femtoscopic contributions decrease, while correlations coming from minijets appear: we can observe the near-side peak for like-sign correlation functions first lowering for the second $p_{T-sum}$ bin, and then rising again for the higher bins.

Moreover, for the highest $p_{T-sum}$ bin the near-side peak is higher for unlike-sign than for like-sign pairs. This is an expected behavior, since due to the local charge conservation minijets are usually composed of both positive and negative particles; therefore, basing on the combinatorics we should have more prominent correlation structure coming from minijets for unlike-sign pairs than for like-sign. This is also observed in Pythia (where no femtoscopic correlations are present) but until now this was not observed for the data because because of the stronger femtoscopic
contribution than the mini-jet contribution at lower energies to the like-sign correlation functions.

\section{Conclusions}
The ALICE experiment registered proton-proton collisions at $\sqrt{s}=0.9\ \rm{TeV}$, $\sqrt{s}=2.76\ \rm{TeV}$ and $\sqrt{s}=7\ \rm{TeV}$ center of mass energies. The collected data allowed us to perform the $\Delta\eta\Delta\phi$ angular correlations analysis at all mentioned energies and as a function of multiplicity and pair transverse momentum.

We observe that femtoscopic and minijet correlations both contribute to the overall correlation 
structure, with minijets correlations being especially prominent for high transverse momenta. We can confirm that the long-range wide structure visible in the femtoscopic analysis presented in \cite{NewFemtoPaper} (strongest for high pair transverse momenta) can originate in the minijet correlations.

\end{document}